\documentclass{article}
\usepackage{graphicx} 
\usepackage{physics}
\usepackage{amsmath,amssymb}
\usepackage{slashed}
\usepackage{bbold}
\usepackage[style=nature, backend=bibtex]{biblatex} 
\usepackage{geometry}
\usepackage{authblk}
\usepackage{abstract}
\usepackage{hyperref}
\usepackage{xurl}
\hypersetup{breaklinks=true}

\def\correspondingauthor{\footnote{cameron.cianci@uconn.edu}}

\title{The Quantum-Extended Church-Turing Thesis in Quantum Field Theory}
\author[1]{Cameron Cianci\correspondingauthor{}}
\affil[1]{Physics Department, University of Connecticut, Storrs, CT 06269, USA}
\date{}

\begin{document}

\maketitle

\begin{abstract}
The quantum-Extended Church-Turing thesis has been explored in many physical theories including general relativity but lacks exploration in quantum field theories such as quantum electrodynamics.  Through construction of a computational model whose gate set mimics the interactions of QED, we demonstrate that one of the defining features of quantum field theory, particle creation and annihilation, is not likely to violate the quantum-Extended Church-Turing thesis.  Through this computational model, it is shown that particle creation is likely only another form of quantum parallelism.  However, whether or not the quantum-Extended Church-Turing thesis will hold for all computational devices in quantum field theories is still not known.  For example, we briefly examine certain interactions in quantum electrodynamics which may create multi-qubit gates.  These gates may have exponential complexity at the cost of being exponentially weak.  This may in turn allow for computational advantage over traditional gate sets such as Clifford+T.
\end{abstract}

\section{The Church-Turing Thesis}
The Church-Turing thesis states that every solvable decision problem can be calculated on a Turing machine \cite{8914d775-0b7b-3433-894d-e3b55f1b0e1a, https://doi.org/10.1112/plms/s2-42.1.230, 88d09d93-3574-3ee8-8658-ea732b3f5a1b}.  This thesis was later extended to complexity theory by proposing that any efficient calculation done by a physical system can be efficiently computed by a Turing machine, where an efficient computation is a computation which can be performed in polynomial time \cite{https://doi.org/10.1112/plms/s2-45.1.161}.  

The extended Church-Turing thesis held until the appearance of Shor's algorithm, which allows quantum computers to factor large numbers in exponentially less time than the best classical algorithm \cite{365700, nielsen_chuang_2010}.  This algorithm implies that certain computational problems may be efficiently solved on a quantum computer but not on a classical Turing machine.  In response to this development, the quantum-Extended Church-Turing thesis was formulated, which proposes that any efficient calculation can be efficiently done by a quantum Turing machine \cite{10.5555/1206629}.

\section{qECT in General Relativity}
Armed with a violation of the extended Church-Turing thesis, physicists sought out new ways to use physics to violate the quantum-Extended Church-Turing thesis.  Many of the most promising approaches to date have been found through utilizing time dilation in general relativity. 

In general relativity, time dilation allows for an observer to initialize a computing machine outside of a black hole, and then cross the event horizon in order to view the result of an infinitely long computation in constant time \cite{NEMETI2006118, 3629bf23-ed55-3b43-9eed-0d6d5819d84b}.  This allows the observer to perform calculations which would otherwise take infinite time on a Turing machine.  As the results of these infinitely long computational problems cannot be computed by a Turing machine, this violates the Church-Turing thesis.  However, Susskind proposed solving this through altering the Church-Turing thesis to require the physical system to be able to communicate with the holographic boundary of space.  This modification prevents the event horizon of a black hole from being used to violate the Church-Turing thesis \cite{susskind2020horizons}.  

Similarly to how the properties of general relativity have been leveraged to test the quantum-Extended Church-Turing thesis, we will investigate if the properties of quantum field theory may allow for computational advantage over traditional quantum computers.

\section{Computing Machines in Quantum Electrodynamics}
Quantum field theory is the most accurate description of the physical world which has been verified through experiment \cite{PhysRevLett.97.030802, PhysRevLett.78.424, PhysRevLett.96.033001}.  The widely successful Standard Model of particle physics is a quantum field theory which encapsulates the strong, weak, and electromagnetic forces, leaving out only gravity \cite{Peskin:1995ev, schwartz_2013, Zee2010}.  Although the study of quantum computing brings computer science a step closer to the fundamental laws of physics, there are still properties of quantum field theory which are not shared by quantum mechanics.  

Foremost of these properties is particle creation and annihilation.  This property allows quantum field theory to create new particles, or new computing subsystems, during a computation.  At a first glance, this property is reminiscent of a nondeterministic Turing machine, which can branch to create new Turing machines and calculate each possible solution to a problem in parallel \cite{sipser13}.  However, if particle creation in quantum field theory indeed behaved this way, it would violate the quantum-Extended Church-Turing thesis, assuming $P \neq NP$.  We will instead demonstrate that this property is likely only another form of quantum parallelism, a property already present in traditional quantum computers.

One of the simplest quantum field theories is Quantum Electrodynamics (QED).  QED only describes electromagnetic interactions, disregarding the Strong and Weak force, as well as gravity.  This theory simply couples a fermionic field to a U(1) gauge boson.  The Lagrangian of quantum electrodynamics is \cite{Peskin:1995ev},

\begin{equation}
    \mathcal{L} = \bar{\psi}(i \gamma^\mu \partial_\mu + m)\psi - \frac{1}{4}F_{\mu\nu}F^{\mu\nu} - ie\bar{\psi}\gamma^\mu A_\mu \psi
\end{equation}

We will use this physical theory to inspire a computational model, constructing a gate set which mimics the interactions of QED.  In this computational model, we will make use of the spin degree of freedom of the photon as a qubit.  Therefore, computational gates which create new photons will also create new qubits.

In quantum field theory, not only can we have a superposition of states, but we can also have a superposition of particle number.  This can be illustrated through the following two states,

\begin{equation}
    \frac{1}{\sqrt{2}}(A_1(x) - A_2(x))\ket{\Omega} \approx \frac{1}{\sqrt{2}}(\ket{0} - \ket{1})
\end{equation}

\begin{equation}
    \frac{1}{2}(A_1(x) - A_2(x)) + \frac{1}{2}(A_1(x)A_1(y) + A_2(x)A_2(y))\ket{\Omega} \approx \frac{1}{2}(\ket{0} - \ket{1}) + \frac{1}{2}(\ket{00} + \ket{11})
\end{equation}

The operator $A_1(x)$ creates a horizontally polarized photon at position $x$, and the operator $A_2(y)$ creates a vertically polarized photon at position $y$.  In this example, horizontally polarized photons are defined as the $\ket{0}$ computational basis state, and vertically polarized photons are defined as the $\ket{1}$ computational basis state.  

Equation 2 shows a photon at position x in a $\ket{-}$ state, a computational state which can be easily obtained in a normal quantum computer. However, equation 3 demonstrates a superposition of particle number, which includes both a single photon in the state $\ket{-}$ in superposition with a bell state $\frac{1}{\sqrt{2}}(\ket{00} + \ket{11})$. Wavefunctions such as this second example allow quantum field theory to realize states in a way which cannot be realized using quantum mechanics, as quantum mechanics preserves particle number \cite{Griffiths2004Introduction}.

\section{Example Computer in QED}
Since calculating the exact interactions and time evolution of an arbitrary state in QED can be incredibly complex, we will instead use a simpler gate based model as a guide.  The gate set of this computational model will mimic the interactions of QED.

The time evolution operator in quantum field theory is found by exponentiating the Hamiltionian to obtain the time evolution operator $U(t)$ \cite{Peskin:1995ev}.

\begin{equation}
    U(t) = e^{-i\mathcal{H}t}  
\end{equation}

Perturbation theory allows us to expand the time evolution operator in orders of the coupling constant $e$.  In this way we can investigate different interactions which take place in QED.  For example, we can see the structure of the following interaction which is present at fourth order in perturbation theory.

\begin{equation}
    \mathcal{O}(e^4) a^\dag_p a^\dag_q a_r a_s \in \frac{(-i\mathcal{H}t)^4}{4!}
\end{equation}

This interaction has the form of a two-qubit gate and is somewhat similar to those used to generate two-qubit gates in superconducting architectures \cite{Magesan_2020, Blais_2021}.  Due to this, we will include a two-qubit gate in the gate set of the QED computational model.  This interaction can be visually represented through the following Feynman diagram, where time is along the x-axis and space is along the y-axis.

\begin{center}
    \includegraphics[scale = 1.0]{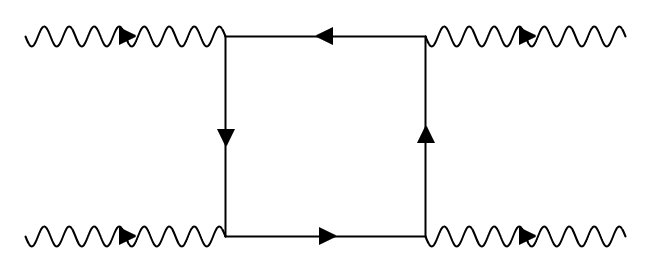}
    
    \textbf{\footnotesize Figure 1. A Feynman diagram whose form suggests a two qubit gate.}
\end{center}

In addition to two-qubit gates, the following interaction is present at sixth order.

\begin{equation}
    \mathcal{O}(e^6) a^\dag_p a^\dag_q a^\dag_r a^\dag_s a_t a_u \in \frac{(-i\mathcal{H}t)^6}{6!}
\end{equation}

Due to this interaction, we will similarly adopt a novel gate into the gate set which can output four qubits from two input qubits, allowing for the computational model to utilize particle creation.  The interaction can be expressed diagramatically using the following Feynman diagram,

\begin{center}
    \includegraphics[scale = .8]{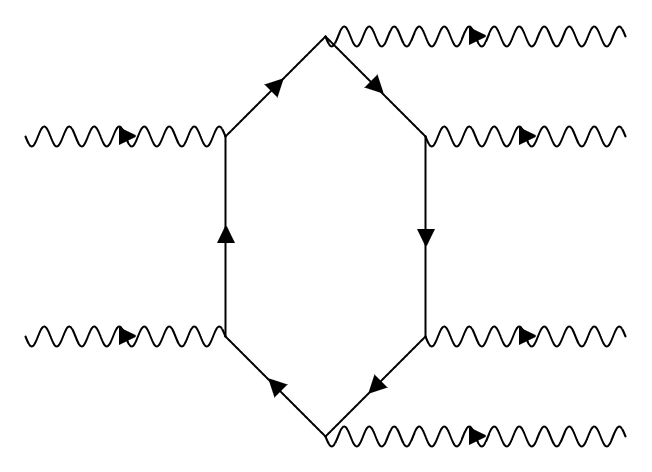}
    
    \textbf{\footnotesize Figure 2. A Feynman diagram in Quantum Electrodynamics which allows for the creation of new qubits.}
\end{center}

As is conventionally done in quantum computing, we can view these gates in a circuit diagram.  For example,

\begin{center}
    \includegraphics[scale = .22]{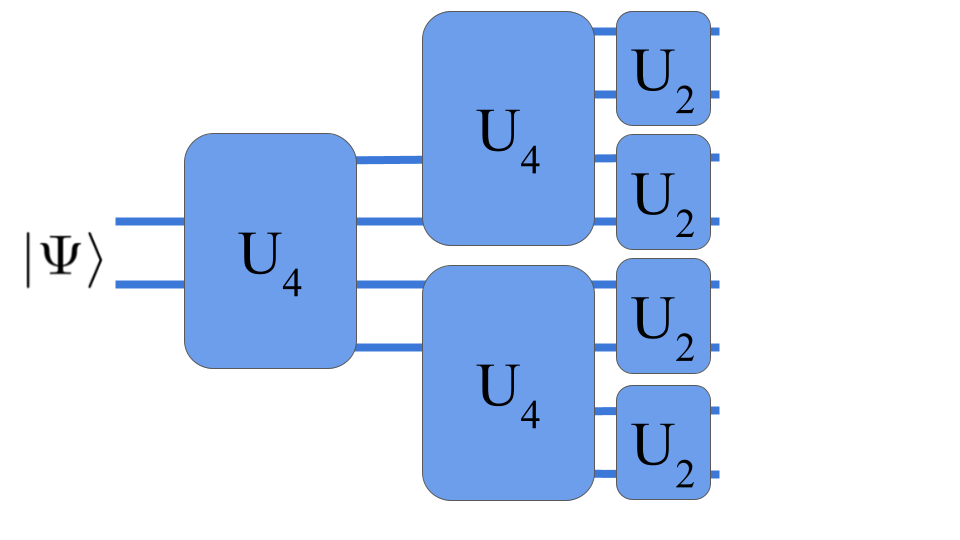}
    
    \textbf{\footnotesize Figure 3. Use of the particle creation gate depicted in a circuit diagram, which may be generated from the interaction in Figure 2.}
\end{center} 

This ability to create new qubits differentiates this computational model from a traditional quantum computer.  This brings about the question, does the addition of this particle creation gate bring computational advantage to devices in quantum field theory?

\section{Equivalence to a Qutrit Quantum Computer}

To answer the previous question, we will now demonstrate that the computation model constructed can be simulated by a quantum computer with an exponential number of ancilla qutrits. Therefore, these particle creation gates would not allow for a violation of the quantum-Extended Church-Turing thesis.

We can demonstrate the equivalence of this QED gate set to a traditional quantum computer as follows.  Each spin degree of freedom in quantum electrodynamics can be translated to 3 states.  The first two states are common to quantum computers, $\ket{0}$, $\ket{1}$, however we also include a third state $\ket{\Omega}$, which indicates that the particle does not exist.

\begin{center}
    \includegraphics[scale = .22]{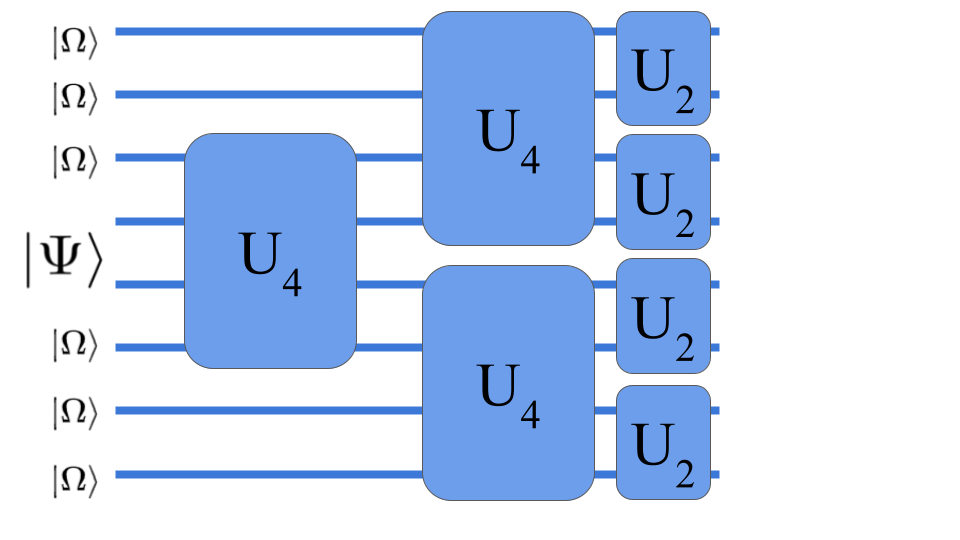}
    
    \textbf{\footnotesize Figure 4. The same circuit as in Figure 3, simulated on a quantum computer with access to ancilla qubits initialized in the $\ket{\Omega}$ state.}
\end{center} 

\newpage

Re-examining the state from Section 3, $\ket{\psi} \approx \frac{1}{\sqrt{2}}(\ket{0} + \ket{1}) + \frac{1}{\sqrt{2}}(\ket{00} + \ket{11})$, we can re-express this state as the qutrit state,

\begin{equation}
     \frac{1}{2}(A_1(x) - A_2(x)) + \frac{1}{2}(A_1(x)A_1(y) + A_2(x)A_2(y))\ket{\Omega} \approx \frac{1}{\sqrt{2}}(\ket{0\Omega} + \ket{1\Omega}) + \frac{1}{\sqrt{2}}(\ket{00} + \ket{11})
\end{equation}

In addition to being able to express a superposition of particle number using qutrits, and since the computing machine is limited to a polynomial amount of time, only an exponential number of particles can be created with the gates included.  In this way, all states which can be reached in exponential time can be expressed with a finite number of qutrits. 

The gate set outlined in the previous sections translate to two-qutrit and four-qutrit gates, which can be efficiently simulated using Clifford+T gates. In this way, any state and any operator which can be efficiently computed in the computational model put forth in Section 4, can be efficiently simulated by a qutrit quantum computer with an exponential number of ancilla qutrits.  Therefore, these particle creation gates would not allow for a violation of the quantum-Extended Church-Turing thesis.

Through understanding that qutrit states in superposition corresponds to a superposition of particle number in the computational model, this computational model suggests that particle creation acts only as another form of quantum parallelism.  This finding demonstrates that particle creation, one of the defining aspects of quantum field theory, will likely not allow for a violation of the quantum-Extended Church-Turing thesis.

\section{Exponentially Small Multi-Qubit Operations}
Although particle creation may not violate the quantum-Extended Church-Turing thesis, this does not imply that other aspects of quantum field theory will not.  There are still other possible mechanisms which may still allow for devices in quantum field theory to gain computational advantage over quantum computers.  

We can discover such an example by investigating the limitations of the constructed gate set.  The four-qutrit particle creation gate shown previously can be simulated efficiently by Clifford+T gates since it has a fixed size.  However, quantum electrodynamics allows for interactions which scale with the system size, and therefore are likely more difficult to simulate with a quantum computer.  These interactions are exponentially suppressed by the coupling constant $e$, suggesting that these interactions may be best modeled by a new class of gates best described as \emph{exponentially suppressed multi-qubit} gates.

A few examples of these high order Feynman diagrams are shown below for a 7-qubit case.

\begin{center}
    \includegraphics[scale = 0.7]{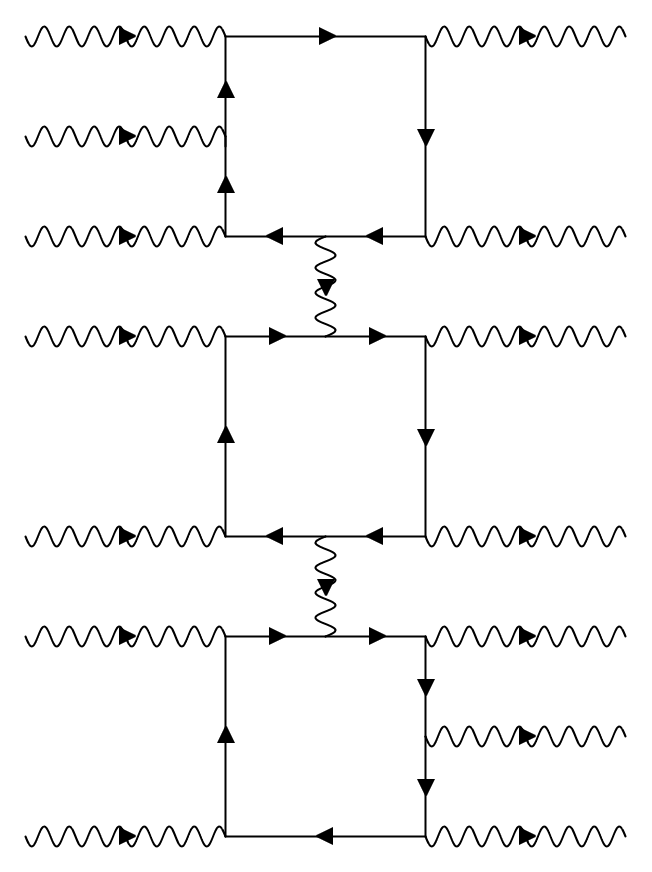}
    \includegraphics[scale = 0.3]{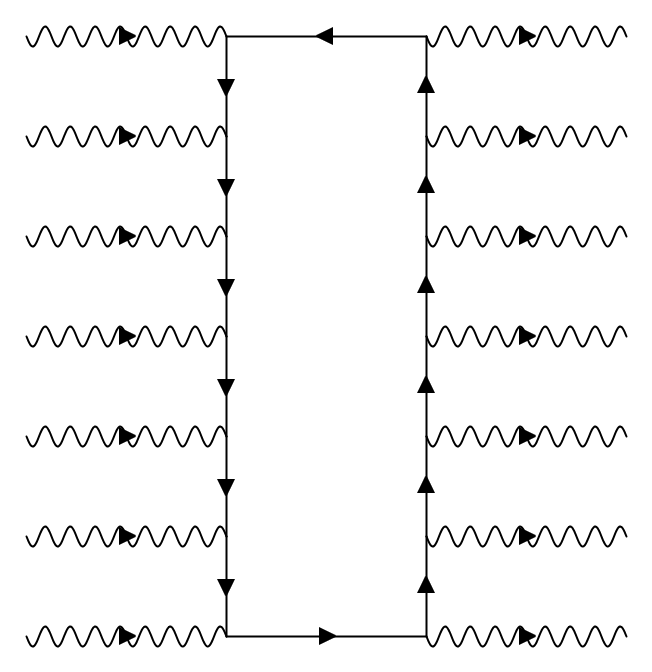}

    \textbf{\footnotesize Figure 5. Example Feynman diagrams which may allow for multi-qubit gates whose magnitude is suppressed exponentially by the coupling constant $\mathcal{O}(e^N$).}
\end{center}

Each vertex suppresses the amplitude of the diagram by a factor of $e$, the coupling constant of electromagnetism.  Therefore, the amplitude of an $N$-qubit operation which leveraged this type of interaction must scale as $e^{N}$.  

As a first approach to modeling these interactions, it may be useful to examine multi-qubit operators $U$ which satisfy the following constraint,

\begin{equation}
    Tr(\mathcal{I} - U) \leq 2^{-n}
\end{equation}

Where $n$ is the number of qubits to which the gate is applied.  These operators may have up to $\mathcal{O}(2^{2n})$ parameters, which makes them exponentially difficult to exactly realize with the traditional Clifford+T gate set.  However, it may be possible to approximate these gates using traditional gate sets.

Examining the successes of quantum computing indicate that access to these gates may provide computational advantage.  The Gottesman-Knill theorem demonstrates that the computational speedup of quantum computers does not arise solely from the properties of entanglement or superposition \cite{gottesman1998heisenberg}.  Furthermore, Shor's algorithm instead suggests that quantum speedups arise from exponentially small finely tuned gates \cite{365700}.  In this way, access to these gates may allow for greater computational power than a traditional gate set such as Clifford+T.

If computational advantage is identified with these gates, then it is important to remember that the quantum-Extended Church-Turing thesis is only violated if these gates can be realized in QED or the Standard Model.  Althought these gates are inspired by exponentially weak mutiparticle interactions in QED, symmetries and other properties of quantum field theories often constrain physically realizable interactions.

\section{Conclusion and Further Directions}
In this paper, we construct a computational model to investigate the computational power of quantum field theories.  Using this model, we demonstrate that particle creation, one of the defining properties of quantum field theory, is not likely to violate the quantum-Extended Church-Turing thesis, as the states and gates of this computational model can be efficiently simulated by a qutrit quantum computer with exponentially many ancilla.

Although particle creation may not allow for computational advantage over a quantum computer, other properties of quantum field theories such as exponentially weak multi-qubit interactions may still allow for computational advantage.

To begin investigating this, it may be useful to examine if the class of operators which satisfy $Tr(\mathcal{I} - U) \approx 2^{-n}$ allow for computational advantage in any known problems, or if these gates can be approximated in polynomial time by traditional gate sets.  Since these operators have exponentially many tunable parameters, it is possible that they may be able to access parts of the state space which would be impossible for normal gate sets to efficiently access.  

Even if computational advantage is identified through this multi-qubit gate set, these gates must still be constructed in a quantum field theory such as QED or the Standard Model in order to claim a violation of the quantum-Extended Church-Turing thesis.  However, seeing as Shor's algorithm finds a computational speedup through exponentially small, finely tuned gates, the computational power of this gate set may be of interest.

\printbibliography

\end{document}